\documentclass{ws-procs10x7}

\begin{document}
\title{Rare Kaon Decay From E949 At BNL: $K^+\rightarrow\pi^+\nu\bar{\nu}$}
\author{
Shaomin Chen\\
$On~behalf~of~E949~Collaboration$}
\address{TRIUMF, 4004 Wesbrook Mall, Vancouver,\\
B.C., V6T 2A3, Canada}
\twocolumn[\maketitle\abstract{
 In the first year of physics run, the E949 experiment at Brookhaven National 
 Laboratory has already collected $1.8\times 10^{12}$ kaons stopping in the 
 target. Additional evidence for the rare charged kaon decay
 $K^+\rightarrow\pi^+\nu\bar{\nu}$ has been observed. Combined
 with previous results from the E787 experiment, the branching ratio
 is measured to be Br($K^+\rightarrow\pi^+\nu\bar{\nu}$)
 =$(1.47^{+1.30}_{-0.89})\times10^{-10}$.}]

\section{Introduction}

The understanding of flavor dynamics is of fundamental importance in elementary
particle physics. The study of $K$ and $B$ decays plays
a crucial role in this respect. In particular, the precise measurement 
of the rare kaon decays $K^+\rightarrow\pi^+\nu\overline\nu$ and
$K^0_L\rightarrow\pi^0\nu\overline\nu$ is regarded as a theoretical
ideal for the understanding of CP violation in the Standard Model (SM).
These modes provide a clean environment for extraction of the CP violation 
related elements in the Cabibbo-Kobayashi-Maskawa quark mixing 
matrix~\cite{rarek}. The unitarity triangle can be completely 
determined by the measurement of these two modes, giving a chance
to confirm the golden relation 
$(sin2\beta)_{\pi\nu\bar{\nu}}=(sin2\beta)_{J/\psi Ks}$,
a key ingredient for the existence of universal
unitarity triangle in the SM and its extensions with minimal flavor
violation~\cite{mfv}. 
Experimentally, only the charged mode studies\cite{e787} 
have achieved  sensitivity comparable to the 
SM prediction\cite{ckmfitter} $(0.67^{+0.28}_{-0.27})\times 10^{-10}$.

In this talk, the first result from E949 experiment at
the Alternating Gradient Synchrotron (AGS) of Brookhaven National
Laboratory (BNL) is presented using a data set with $1.8\times10^{12}$ 
stopping charged kaons.

\section{Backgrounds in E949}

Like its predecessor E787, the E949 detector is designed to clearly identify
$K^+\rightarrow\pi^+\nu\bar{\nu}$, with a charged kaon 
decaying to a charged pion in the momentum region $221<P<229$ MeV/$c$ with 
no other associated observable products. Potential backgrounds come from
$K^+\rightarrow\mu^+\nu_\mu(\gamma)$ (due to $\mu^+$ misidentified as
$\pi^+$ and mis-measured kinematics or missed photon),
$K^+\rightarrow\pi^+\pi^0$ (due to missed photon and mis-measured
kinematics), beam backgrounds (due to incoming $\pi^+$
mis-identified as $K^+$ or $\pi^+$ faking $K^+$ decay at rest
or $K^+$ decay in flight or two incoming beam particles), 
or charge exchange background (due to $K^+ n\rightarrow K^0 p$,
$K^0_L\rightarrow\pi^+ l^-\nu_l$ with $l^-$ undetected). 
Because of the huge background (see Table~\ref{bkg_rej}), 
the experiment is rather challenging, requiring both powerful and 
redundant background rejection tools. Most of the details on how the 
E949 detector suppresses the background processes are the same as 
E787~\cite{e787,chen}. Table~\ref{bkg_rej} lists the basic tools 
for rejecting the backgrounds that can look like 
$K^+\rightarrow\pi^+\nu\bar{\nu}$.  
\begin{table*}
\caption{Comparison of signal and background yields normalized to single 
signal event production and the selection criteria (cuts)
for suppressing the backgrounds. 
The kinematic cuts are defined as those using
the momentum, energy and range in scintillator. PID is defined
as either pion identification with TD (see text) or kaon 
identification using the B4 dE/dx. PV is for
the photon veto and the timing cuts are for the delayed coincidence cuts.
}
\begin{tabular}{|l|l|l|l|l|l|}\hline
Cut    &Events & Kinematic cuts & PID & PV & Timing cuts \\ \hline 
$K^+\rightarrow\mu^+\nu_\mu$       & 6343000000      
       &Yes&Yes&$-$&$-$\\ \hline
$K^+\rightarrow\pi^+\pi^0$         & 2113000000     
        &Yes&$-$&Yes&$-$\\ \hline
$K^+\rightarrow\mu^+\nu_\mu\gamma$ &   55000000    
         &Yes&Yes&Yes&$-$\\ \hline 
Beam background      &   25000000                  
         &$-$&Yes&$-$&Yes\\ \hline
$K^+n\rightarrow K^0p,~K^0\rightarrow\pi^+l^-\nu$&      46000
&$-$&$-$&Yes&Yes \\ \hline
$K^+\rightarrow\pi^+\nu\bar{\nu}$               &          1
&Yes&Yes&Yes&Yes\\ \hline
\end{tabular}
\label{bkg_rej}
\end{table*}

\section{Detector upgrades}

E949 contains several various  upgrades to the E787 detector.
Here only the most important changes are described.
For the beam counters, the B4 hodoscope in front of the target fibers 
were redesigned to improve the position measurement of the incoming beam, 
leading to a factor of $\sqrt{2}$ improvement. To gain more rejection of
photons, E949 increased the thickness and solid angle of the active 
material for detecting photon activity by adding a barrel veto 
liner outside the 19th layer of scintillator, where two layers of 
scintillator were previously installed . The photon veto 
rejection (PV) was measured to be a factor 
of two greater than E787 at the same acceptance.
To improve the $z$-position measurement of charged tracks and 
increase the kinematic rejection against the
muon background, a dual threshold discriminator 
system was applied to the straw chambers embedded in the 19-layer
scintillator stack, significantly improving their performance.
Following these upgrades, higher intensity operation was improved, with
the momentum, range and energy resolutions of
$\Delta P/P\sim 1.1\%$ at 205 MeV/$c$,
$\Delta R/R\sim 2.8\%$ at 30.5 cm and 
$\Delta E/\sqrt{E(\mbox{GeV})}\sim 0.9\%$ at 108.5 MeV, respectively. 

\section{Data analysis}

This study of $K^+\rightarrow\pi^+\nu\bar{\nu}$ adopted the technique of 
blind analysis, in which a pre-defined signal region was not examined until
all background rates had been estimated and verified. 
The cuts were designed from a 
study of background samples to avoid bias. Before looking into
the signal region (``opening the box"), background levels were estimated 
using the knowledge outside the signal region. Cuts were designed  
to bring the background down to the $\leq$0.5 event level.  

Although each cut setting should be based on sufficient statistics,
bias can still occur and may cause a problem
in background estimate. In order to have unbiased cuts, the full data 
set was randomly divided into 1/3 and 2/3 parts. Cut tuning 
was conducted only on the 1/3 sample, in which three specific sub-data 
groups were also selected to simplify the analysis. By applying both loose beam
background cuts and photon veto, we skimmed out a sample dominated by
$K^+\rightarrow\mu^+\nu_{\mu}(\gamma)$ background for 
tuning the pion particle ID which involves positive identification 
of the $\pi\rightarrow\mu\rightarrow e$ decay chain in the range stack using 
the transient digitizers (TD) 
and the kinematic cuts. The sample with $K^+\rightarrow\pi^+\pi^0$ 
dominant background was selected by using both loose
beam background cuts and loose TD cuts. Finally, applying 
both loose TD cuts and photon veto, we obtained a sample
dominated by the beam background. 

As detailed elsewhere~\cite{e787,chen}, two uncorrelated cuts giving 
large background rejection were selected to perform a
so-called bifurcated analysis on a single background. 
By sequentially inverting one of the cuts and counting the events 
observed, we estimated the background inside the signal region 
when both cuts were applied. Table~\ref{bkg_rej} gives the two uncorrelated
cuts for each background. 

\begin{figure}[ht]
\epsfxsize200pt
\figurebox{120pt}{160pt}{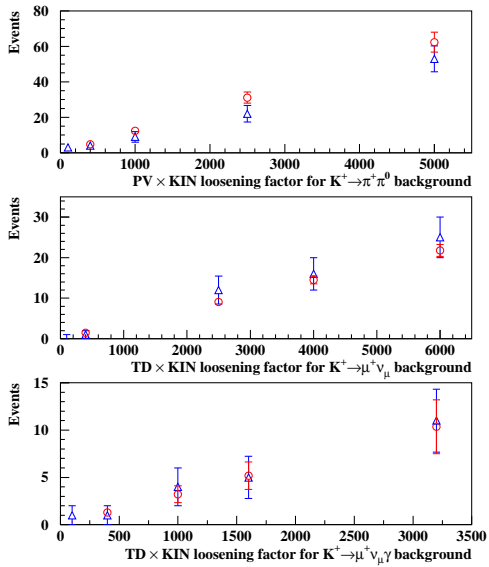}
\caption{The outside-the-box study backgrounds 
$K^+\rightarrow\pi^+\pi^0$ and
$K^+\rightarrow\mu^+\nu_\mu(\gamma)$ $K^+\rightarrow\pi^+\pi^0$.
The open circles are for the predicted backgrounds and the  
open triangles are for the observed backgrounds.}
\label{outside}
\end{figure}

To check if the cut pairs are uncorrelated, the regions near, 
but outside, the signal region (outside the box) was 
examined by loosening both cuts and checking if the observed number
of events was consistent with the estimated background level.
As shown in Figure~\ref{outside},
the observation agrees with the prediction as the cuts are loosened.
Using a one-parameter fit to the ratio of $N_{obs}/N_{pred}$, 
we obtained $0.85^{+0.12}_{-0.11}$, $1.15^{+0.25}_{-0.21}$ and
$1.06^{+0.35}_{-0.29}$ for $K^+\rightarrow\pi^+\pi^0$,
$K^+\rightarrow\mu^+\nu_\mu$ and $K^+\rightarrow\mu^+\nu_\mu\gamma$,
respectively. These results are consistent with the expected 1.0.
The errors on the constant terms are assigned as systematic uncertainties
and are used in the final determination of the branching ratio. 

We also considered  the possibility  that other backgrounds
could sneak into the signal region through possible unknown loopholes 
in our procedures. Limitations on such effects came from the-outside-box 
study which used data very close to the signal box. We carefully examined 
the background events which failed only a single cut.

The final background estimates are given in Table~\ref{total_bkg}. 
The background estimates for E787 are given for comparison. Since 
we had much confidence on the cuts used in control of
the $K^+\rightarrow\pi^+\pi^0$ background, we loosened the cuts
in order to look into a larger signal region.
Consequently, the $K^+\rightarrow\pi^+\pi^0$ background level increases
in E949.
\begin{table}
\caption{Total stopping kaons and background estimates 
in E949 and E787. The charge exchange background 
estimate is from Monte Carlo. The kaon regeneration rate and beam 
profile are from measurement for the process of 
$K^+n\rightarrow K^0_Sp, K^0_S\rightarrow\pi^+\pi^-$.
Errors are statistical only.}
\begin{tabular}{|l|l|l|}\hline
Item                  & E949               & E787\\ \hline
$N_K$                 & $1.8\times10^{12}$ & $5.9\times10^{12}$ \\ \hline 
$\pi^+\pi^0$          & $0.216\pm0.023$    & $0.034\pm0.007$ \\ \hline
$\mu^+\nu_\mu(\gamma)$& $0.068\pm0.011$    & $0.062\pm0.045$ \\ \hline
Beam bkg              & $0.009\pm0.003$    & $0.025\pm0.016$ \\ \hline
CEX                   & $0.005\pm0.001$    & $0.025\pm0.008$ \\ \hline
Total bkg             & $0.298\pm0.026$    & $0.146\pm0.049$ \\ \hline
\end{tabular}
\label{total_bkg}
\end{table}

\section{Branching ratio measurement}

\begin{figure}[ht]
\epsfxsize200pt
\figurebox{120pt}{160pt}{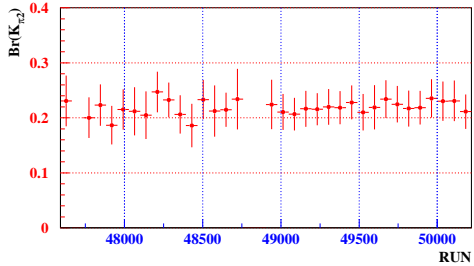}
\caption{Measurement of the $K^+\rightarrow\pi^+\pi^0$ branching ratio 
as a function of run number in E949.}
\label{brkpi2}
\end{figure}

In E949, the final acceptance for $K^+\rightarrow\pi^+\nu\bar{\nu}$ 
was estimated to be $0.0022\pm0.0002$, slightly higher than 
$0.0020\pm0.0002$ of E787. The final acceptance was checked 
with a measurement of the $K^+\rightarrow\pi^+\pi^0$  branching ratio, 
which was measured to be $0.219\pm0.005$, in agreement with the 
PDG value\cite{pdg} $0.211\pm0.001$ within two standard deviations.
The stability of this branching ratio measurement throughout the run 
is shown in Figure~\ref{brkpi2}. 

\begin{figure}[ht]
\epsfxsize200pt
\figurebox{120pt}{160pt}{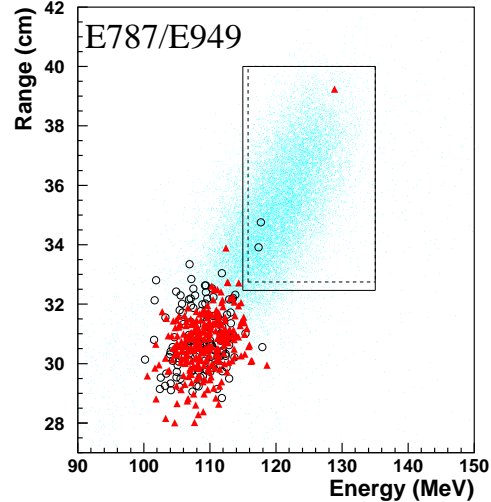}
\caption{Range versus kinetic energy distribution with all other cuts 
applied. The circles, triangles and dots represent E787, E949 data 
and the simulation of events from $K^+\rightarrow\pi^+\nu\bar{\nu}$ 
decay, respectively. The signal box used in E787 and E949 are 
indicated by the dashed and the solid line separately. 
}
\label{box}
\end{figure}

The signal box was opened after the background study and 
acceptance estimate were completed. One candidate was observed inside the box. 
Figure~\ref{box} shows the range versus kinetic energy for the events
surviving all the selection criteria. The events outside this box are
from the $K^+\rightarrow\pi^+\pi^0$ background due to photons escaping
detection. 

Because the background level in Table~\ref{total_bkg} was for the total 
number inside the signal box, knowing the fact that the background 
distribution inside the box is not uniform, we used
the background distributions to evaluate the background 
probability for a candidate event and  to improve the estimate of the 
branching ratio~\cite{e787}. It was found that the E949 candidate event 
was closer to the muon background region than the previous E787 events,
reducing its contribution to the final branching ratio
measured to be Br($K^+\rightarrow\pi^+\nu\bar{\nu}$)
 =$(1.47^{+1.30}_{-0.89})\times10^{-10}$. 

\section{Conclusion}

The first physics run of E949 has been successfully carried out
with a factor of two higher beam intensity than E787.
One new candidate event of $K^+\rightarrow\pi^+\nu\bar{\nu}$
was observed. The central value of the branching ratio is 
approximately twice the SM prediction, and the 1$\sigma$ confidence 
interval includes the SM and up to 4.0 times the SM.
Completion of the remaining 80\% of running time
already approved by the U.S. Department of Energy's Office of High
Energy Physics should indicate whether new physics is increasing this
branching ratio.

\section*{Acknowledgments}

I am indebted to my colleagues of the E949 Collaboration, 
in particular to D. Bryman, D. Jaffe and S. Kettell for their
discussion and comments.

\end{document}